\input harvmac

\input amssym

\def\unit{\relax{\rm 1\kern-.26em I}}
\def\nada{\relax{\rm 0\kern-.30em l}}
\def\tilde{\widetilde}

\def\alphadot{{\dot \alpha}}
\def\betadot{{\dot\beta}}



\def\det{{\rm det}}

\noblackbox
\def\IL{\relax{\rm I\kern-.18em L}}
\def\IH{\relax{\rm I\kern-.18em H}}
\def\IR{\relax{\rm I\kern-.18em R}}
\def\IP{\relax{\rm I\kern-.18em P}}
\def\IC{\relax\hbox{$\inbar\kern-.3em{\rm C}$}}
\def\IZ{\relax\ifmmode\mathchoice
{\hbox{\cmss Z\kern-.4em Z}}{\hbox{\cmss Z\kern-.4em Z}} {\lower.9pt\hbox{\cmsss Z\kern-.4em Z}}
{\lower1.2pt\hbox{\cmsss Z\kern-.4em Z}}\else{\cmss Z\kern-.4em Z}\fi}
\def\CM {{\cal M}}
\def\CN {{\cal N}}
\def\CR {{\cal R}}

\def\CJ {{\cal J}}

\def\CL {{\cal L}}

\def\CS {{\cal S}}
\def\CA{{\cal A}}

\def\CM {{\cal M}}
\def\CN {{\cal N}}

\def\CS {{\cal S }}

\def\det{{\rm det}}

\font\manual=manfnt \def\dbend{\lower3.5pt\hbox{\manual\char127}}

\def\IZ{\relax\ifmmode\mathchoice
{\hbox{\cmss Z\kern-.4em Z}}{\hbox{\cmss Z\kern-.4em Z}} {\lower.9pt\hbox{\cmsss Z\kern-.4em Z}}
{\lower1.2pt\hbox{\cmsss Z\kern-.4em Z}}\else{\cmss Z\kern-.4em Z}\fi}

\def\bar{\overline}
\def\CS{{\cal S}}

\def\pa{\partial}

\def\rt2{\sqrt{2}}
\def\irt2{{1\over\sqrt{2}}}

\def\hat{\widehat}
\def\slashchar#1{\setbox0=\hbox{$#1$}           
   \dimen0=\wd0                                 
   \setbox1=\hbox{/} \dimen1=\wd1               
   \ifdim\dimen0>\dimen1                        
      \rlap{\hbox to \dimen0{\hfil/\hfil}}      
      #1                                        
   \else                                        
      \rlap{\hbox to \dimen1{\hfil$#1$\hfil}}   
      /                                         
   \fi}

\def\foursqr#1#2{{\vcenter{\vbox{
    \hrule height.#2pt
    \hbox{\vrule width.#2pt height#1pt \kern#1pt
    \vrule width.#2pt}
    \hrule height.#2pt
    \hrule height.#2pt
    \hbox{\vrule width.#2pt height#1pt \kern#1pt
    \vrule width.#2pt}
    \hrule height.#2pt
        \hrule height.#2pt
    \hbox{\vrule width.#2pt height#1pt \kern#1pt
    \vrule width.#2pt}
    \hrule height.#2pt
        \hrule height.#2pt
    \hbox{\vrule width.#2pt height#1pt \kern#1pt
    \vrule width.#2pt}
    \hrule height.#2pt}}}}
\def\psqr#1#2{{\vcenter{\vbox{\hrule height.#2pt
    \hbox{\vrule width.#2pt height#1pt \kern#1pt
    \vrule width.#2pt}
    \hrule height.#2pt \hrule height.#2pt
    \hbox{\vrule width.#2pt height#1pt \kern#1pt
    \vrule width.#2pt}
    \hrule height.#2pt}}}}
\def\sqr#1#2{{\vcenter{\vbox{\hrule height.#2pt
    \hbox{\vrule width.#2pt height#1pt \kern#1pt
    \vrule width.#2pt}
    \hrule height.#2pt}}}}

\def\figin{\epsfcheck\figin}\def\figins{\epsfcheck\figins}
\def\epsfcheck{\ifx\epsfbox\UnDeFiNeD
\message{(NO epsf.tex, FIGURES WILL BE IGNORED)}
\gdef\figin##1{\vskip2in}\gdef\figins##1{\hskip.5in}
\else\message{(FIGURES WILL BE INCLUDED)}%
\gdef\figin##1{##1}\gdef\figins##1{##1}\fi}
\def\DefWarn#1{}
\def\figinsert{\goodbreak\midinsert}
\def\ifig#1#2#3{\DefWarn#1\xdef#1{fig.~\the\figno}
\writedef{#1\leftbracket fig.\noexpand~\the\figno}%
\figinsert\figin{\centerline{#3}}\medskip\centerline{\vbox{\baselineskip12pt \advance\hsize by
-1truein\noindent\footnotefont{\bf Fig.~\the\figno:\ } \it#2}}
\bigskip\endinsert\global\advance\figno by1}

\lref\KuzenkoYM{
  S.~M.~Kuzenko,
  ``The Fayet-Iliopoulos term and nonlinear self-duality,''
  arXiv:0911.5190 [hep-th].
}

\lref\ClarkJX{
  T.~E.~Clark, O.~Piguet and K.~Sibold,
  ``Supercurrents, Renormalization And Anomalies,''
  Nucl.\ Phys.\  B {\bf 143}, 445 (1978).
}

\lref\KomargodskiPC{
  Z.~Komargodski and N.~Seiberg,
  ``Comments on the Fayet-Iliopoulos Term in Field Theory and Supergravity,''
  JHEP {\bf 0906}, 007 (2009)
  [arXiv:0904.1159 [hep-th]].
}

\lref\AffleckVC{
  I.~Affleck, M.~Dine and N.~Seiberg,
  ``Dynamical Supersymmetry Breaking In Chiral Theories,''
  Phys.\ Lett.\ B {\bf 137}, 187 (1984).
}

\lref\GatesNR{
  S.~J.~Gates, M.~T.~Grisaru, M.~Rocek and W.~Siegel,
  ``Superspace, or one thousand and one lessons in supersymmetry,''
  Front.\ Phys.\  {\bf 58}, 1 (1983)
  [arXiv:hep-th/0108200].
}

\lref\RocekPC{M.~Rocek, Private communication.}

\lref\SeibergBZ{
  N.~Seiberg,
  ``Exact results on the space of vacua of four-dimensional SUSY gauge
  theories,''
  Phys.\ Rev.\ D {\bf 49}, 6857 (1994)
  [arXiv:hep-th/9402044].
  }

\lref\KalloshVE{
  R.~Kallosh, L.~Kofman, A.~D.~Linde and A.~Van Proeyen,
  ``Superconformal symmetry, supergravity and cosmology,''
  Class.\ Quant.\ Grav.\  {\bf 17}, 4269 (2000)
  [Erratum-ibid.\  {\bf 21}, 5017 (2004)]
  [arXiv:hep-th/0006179].
}

\lref\DvaliZH{
  G.~Dvali, R.~Kallosh and A.~Van Proeyen,
  ``D-term strings,''
  JHEP {\bf 0401}, 035 (2004)
  [arXiv:hep-th/0312005].
}

\lref\FischlerZK{
  W.~Fischler, H.~P.~Nilles, J.~Polchinski, S.~Raby and L.~Susskind,
  ``Vanishing Renormalization Of The D Term In Supersymmetric U(1) Theories,''
  Phys.\ Rev.\ Lett.\  {\bf 47}, 757 (1981).
}

\lref\WittenNF{
  E.~Witten,
  ``Dynamical Breaking Of Supersymmetry,''
  Nucl.\ Phys.\  B {\bf 188}, 513 (1981).
}

\lref\NemeschanskyYX{
  D.~Nemeschansky and A.~Sen,
  ``Conformal Invariance of Supersymmetric Sigma Models on Calabi-Yau
  Manifolds,''
  Phys.\ Lett.\  B {\bf 178}, 365 (1986).
}

\lref\EW{E.~Witten, unpublished.}

\lref\KS{Z.~Komargodski and N.~Seiberg, to appear.}

\lref\ElvangJK{
  H.~Elvang, D.~Z.~Freedman and B.~Kors,
  ``Anomaly cancellation in supergravity with Fayet-Iliopoulos couplings,''
  JHEP {\bf 0611}, 068 (2006)
  [arXiv:hep-th/0606012].
}

\lref\ShifmanZI{
  M.~A.~Shifman and A.~I.~Vainshtein,
  ``Solution of the Anomaly Puzzle in SUSY Gauge Theories and the Wilson
  Operator Expansion,''
  Nucl.\ Phys.\  B {\bf 277}, 456 (1986)
  [Sov.\ Phys.\ JETP {\bf 64}, 428 (1986\ ZETFA,91,723-744.1986)].
}

\lref\BarbieriAC{
  R.~Barbieri, S.~Ferrara, D.~V.~Nanopoulos and K.~S.~Stelle,
  ``Supergravity, R Invariance And Spontaneous Supersymmetry Breaking,''
  Phys.\ Lett.\  B {\bf 113}, 219 (1982).
}

\lref\SeibergVC{
  N.~Seiberg,
  ``Naturalness Versus Supersymmetric Non-renormalization Theorems,''
  Phys.\ Lett.\  B {\bf 318}, 469 (1993)
  [arXiv:hep-ph/9309335].
}

\lref\DineXK{
  M.~Dine, N.~Seiberg and E.~Witten,
  ``Fayet-Iliopoulos Terms in String Theory,''
  Nucl.\ Phys.\  B {\bf 289}, 589 (1987).
}

\lref\FreedmanUK{
  D.~Z.~Freedman,
  ``Supergravity With Axial Gauge Invariance,''
  Phys.\ Rev.\  D {\bf 15}, 1173 (1977).
}

\lref\DasPU{
  A.~Das, M.~Fischler and M.~Rocek,
  ``Superhiggs Effect In A New Class Of Scalar Models And A Model Of Super
  QED,''
  Phys.\ Rev.\  D {\bf 16}, 3427 (1977).
}

\lref\BinetruyHH{
  P.~Binetruy, G.~Dvali, R.~Kallosh and A.~Van Proeyen,
  ``Fayet-Iliopoulos terms in supergravity and cosmology,''
  Class.\ Quant.\ Grav.\  {\bf 21}, 3137 (2004)
  [arXiv:hep-th/0402046].
}

\lref\FerraraDH{
  S.~Ferrara, L.~Girardello, T.~Kugo and A.~Van Proeyen,
  ``Relation Between Different Auxiliary Field Formulations Of N=1 Supergravity
  Coupled To Matter,''
  Nucl.\ Phys.\  B {\bf 223}, 191 (1983).
}

\lref\WeinbergUV{
  S.~Weinberg,
  ``Non-renormalization theorems in non-renormalizable theories,''
  Phys.\ Rev.\ Lett.\  {\bf 80}, 3702 (1998)
  [arXiv:hep-th/9803099].
}

\lref\FerraraPZ{
  S.~Ferrara and B.~Zumino,
  ``Transformation Properties Of The Supercurrent,''
  Nucl.\ Phys.\  B {\bf 87}, 207 (1975).
}

\lref\DineTA{
  M.~Dine,
  ``Fields, Strings and Duality: TASI 96,''
 eds. C.~Efthimiou and B. Greene (World Scientific, Singapore, 1997).
   }

\lref\WittenBZ{
  E.~Witten,
  ``New Issues In Manifolds Of SU(3) Holonomy,''
  Nucl.\ Phys.\  B {\bf 268}, 79 (1986).
}

\lref\SeibergVC{
  N.~Seiberg,
 ``Naturalness Versus Supersymmetric Non-renormalization Theorems,''
  Phys.\ Lett.\  B {\bf 318}, 469 (1993)
  [arXiv:hep-ph/9309335].
}

\lref\ORaifeartaighPR{
  L.~O'Raifeartaigh,
  ``Spontaneous Symmetry Breaking For Chiral Scalar Superfields,''
  Nucl.\ Phys.\  B {\bf 96}, 331 (1975).
}

\lref\FayetJB{
  P.~Fayet and J.~Iliopoulos,
  ``Spontaneously Broken Supergauge Symmetries and Goldstone Spinors,''
  Phys.\ Lett.\  B {\bf 51}, 461 (1974).
}

\lref\GreenSG{
  M.~B.~Green and J.~H.~Schwarz,
  ``Anomaly Cancellation In Supersymmetric D=10 Gauge Theory And Superstring
  Theory,''
  Phys.\ Lett.\  B {\bf 149}, 117 (1984).
}

\lref\ChamseddineGB{
  A.~H.~Chamseddine and H.~K.~Dreiner,
  ``Anomaly Free Gauged R Symmetry In Local Supersymmetry,''
  Nucl.\ Phys.\  B {\bf 458}, 65 (1996)
  [arXiv:hep-ph/9504337].
}

\lref\FerraraMV{
  S.~Ferrara and B.~Zumino,
  ``Structure Of Conformal Supergravity,''
  Nucl.\ Phys.\  B {\bf 134}, 301 (1978).
}

\lref\CastanoCI{
  D.~J.~Castano, D.~Z.~Freedman and C.~Manuel,
  ``Consequences of supergravity with gauged U(1)-R symmetry,''
  Nucl.\ Phys.\  B {\bf 461}, 50 (1996)
  [arXiv:hep-ph/9507397].
}

\lref\WittenHU{
  E.~Witten and J.~Bagger,
  ``Quantization Of Newton's Constant In Certain Supergravity Theories,''
  Phys.\ Lett.\  B {\bf 115}, 202 (1982).
}

\lref\BaggerFN{
  J.~Bagger and E.~Witten,
  ``The Gauge Invariant Supersymmetric Nonlinear Sigma Model,''
  Phys.\ Lett.\  B {\bf 118}, 103 (1982).
}

\lref\GirardelloWZ{
  L.~Girardello and M.~T.~Grisaru,
  ``Soft Breaking Of Supersymmetry,''
  Nucl.\ Phys.\  B {\bf 194}, 65 (1982).
}

\lref\AffleckVC{
  I.~Affleck, M.~Dine and N.~Seiberg,
  ``Dynamical Supersymmetry Breaking In Chiral Theories,''
  Phys.\ Lett.\  B {\bf 137}, 187 (1984).
}

\lref\WessCP{
  J.~Wess and J.~Bagger,
  ``Supersymmetry and supergravity,''
{\it  Princeton, USA: Univ. Pr. (1992) 259 p}
}

\lref\BrignoleSK{
  A.~Brignole, F.~Feruglio and F.~Zwirner,
  ``Signals of a superlight gravitino at e+ e- colliders when the other
  superparticles are heavy,''
  Nucl.\ Phys.\  B {\bf 516}, 13 (1998)
  [Erratum-ibid.\  B {\bf 555}, 653 (1999)]
  [arXiv:hep-ph/9711516].
}

\lref\DineXI{
  M.~Dine, N.~Seiberg and S.~Thomas,
  ``Higgs Physics as a Window Beyond the MSSM (BMSSM),''
  Phys.\ Rev.\  D {\bf 76}, 095004 (2007)
  [arXiv:0707.0005 [hep-ph]].
}

\lref\KomargodskiPC{
  Z.~Komargodski and N.~Seiberg,
  ``Comments on the Fayet-Iliopoulos Term in Field Theory and Supergravity,''
  JHEP {\bf 0906}, 007 (2009)
  [arXiv:0904.1159 [hep-th]].
}

\lref\AffleckVC{
  I.~Affleck, M.~Dine and N.~Seiberg,
  ``Dynamical Supersymmetry Breaking In Chiral Theories,''
  Phys.\ Lett.\  B {\bf 137}, 187 (1984).
}

\lref\ClarkBG{
  T.~E.~Clark and S.~T.~Love,
  ``The Supercurrent in supersymmetric field theories,''
  Int.\ J.\ Mod.\ Phys.\  A {\bf 11}, 2807 (1996)
  [arXiv:hep-th/9506145].
}

\lref\BrignoleFN{
  A.~Brignole, F.~Feruglio and F.~Zwirner,
  ``Aspects of spontaneously broken N = 1 global supersymmetry in the  presence
  of gauge interactions,''
  Nucl.\ Phys.\  B {\bf 501}, 332 (1997)
  [arXiv:hep-ph/9703286].
}

\lref\AffleckMF{
  I.~Affleck, M.~Dine and N.~Seiberg,
  ``Exponential Hierarchy From Dynamical Supersymmetry Breaking,''
  Phys.\ Lett.\  B {\bf 140}, 59 (1984).
}

\lref\AffleckXZ{
  I.~Affleck, M.~Dine and N.~Seiberg,
  ``Dynamical Supersymmetry Breaking In Four-Dimensions And Its
  Phenomenological Implications,''
  Nucl.\ Phys.\  B {\bf 256}, 557 (1985).
}

\lref\WeinbergKR{
  S.~Weinberg,
  ``The quantum theory of fields. Vol. 2: Modern applications,''
{\it  Cambridge, UK: Univ. Pr. (1996) 489 p}
}

\lref\UematsuRJ{
  T.~Uematsu and C.~K.~Zachos,
  ``Structure Of Phenomenological Lagrangians For Broken Supersymmetry,''
  Nucl.\ Phys.\  B {\bf 201}, 250 (1982).
}

\lref\DineII{
  M.~Dine, R.~Kitano, A.~Morisse and Y.~Shirman,
  ``Moduli decays and gravitinos,''
  Phys.\ Rev.\  D {\bf 73}, 123518 (2006)
  [arXiv:hep-ph/0604140].
}

\lref\WittenYC{
  E.~Witten,
  ``Phases of N = 2 theories in two dimensions,''
  Nucl.\ Phys.\  B {\bf 403}, 159 (1993)
  [arXiv:hep-th/9301042].
}

\lref\KachruEM{
  S.~Kachru, J.~McGreevy and P.~Svrcek,
  ``Bounds on masses of bulk fields in string compactifications,''
  JHEP {\bf 0604}, 023 (2006)
  [arXiv:hep-th/0601111].
}

\lref\KachruXP{
  S.~Kachru, L.~McAllister and R.~Sundrum,
  ``Sequestering in string theory,''
  JHEP {\bf 0710}, 013 (2007)
  [arXiv:hep-th/0703105].
}

\lref\RandallUK{
  L.~Randall and R.~Sundrum,
  ``Out of this world supersymmetry breaking,''
  Nucl.\ Phys.\  B {\bf 557}, 79 (1999)
  [arXiv:hep-th/9810155].
}

\lref\DienesTD{
  K.~R.~Dienes and B.~Thomas,
  ``On the Inconsistency of Fayet-Iliopoulos Terms in Supergravity Theories,''
  arXiv:0911.0677 [hep-th].
}

\lref\ArkaniHamedMJ{
  N.~Arkani-Hamed and H.~Murayama,
  ``Holomorphy, rescaling anomalies and exact beta functions in  supersymmetric
  gauge theories,''
  JHEP {\bf 0006}, 030 (2000)
  [arXiv:hep-th/9707133].
}

\lref\GrisaruYK{
  M.~T.~Grisaru, B.~Milewski and D.~Zanon,
  ``Supercurrents, Anomalies And The Adler-Bardeen Theorem,''
  Phys.\ Lett.\  B {\bf 157}, 174 (1985).
}

\lref\GrisaruIK{
  M.~T.~Grisaru, B.~Milewski and D.~Zanon,
  ``The Supercurrent And The Adler-Bardeen Theorem,''
  Nucl.\ Phys.\  B {\bf 266}, 589 (1986).
}

\lref\ShifmanZI{
  M.~A.~Shifman and A.~I.~Vainshtein,
 ``Solution of the Anomaly Puzzle in SUSY Gauge Theories and the Wilson
 Operator Expansion,''
  Nucl.\ Phys.\  B {\bf 277}, 456 (1986)
  [Sov.\ Phys.\ JETP {\bf 64}, 428 (1986\ ZETFA,91,723-744.1986)].
}

\lref\WeinbergCR{
  S.~Weinberg,
  ``The quantum theory of fields.  Vol. 3: Supersymmetry,''
{\it  Cambridge, UK: Univ. Pr. (2000) 419 p} }

\lref\StelleYE{
  K.~S.~Stelle and P.~C.~West,
  ``Minimal Auxiliary Fields For Supergravity,''
  Phys.\ Lett.\  B {\bf 74}, 330 (1978).
}

\lref\FerraraEM{
  S.~Ferrara and P.~van Nieuwenhuizen,
  ``The Auxiliary Fields Of Supergravity,''
  Phys.\ Lett.\  B {\bf 74}, 333 (1978).
}

\lref\SohniusTP{
  M.~F.~Sohnius and P.~C.~West,
  ``An Alternative Minimal Off-Shell Version Of N=1 Supergravity,''
  Phys.\ Lett.\  B {\bf 105}, 353 (1981).
}

\lref\GirardiVQ{
  G.~Girardi, R.~Grimm, M.~Muller and J.~Wess,
  ``Antisymmetric Tensor Gauge Potential In Curved Superspace And A (16+16)
  Supergravity Multiplet,''
  Phys.\ Lett.\  B {\bf 147}, 81 (1984).
}

\lref\LangXK{
  W.~Lang, J.~Louis and B.~A.~Ovrut,
  ``(16+16) Supergravity Coupled To Matter: The Low-Energy Limit Of The
  Superstring,''
  Phys.\ Lett.\  B {\bf 158}, 40 (1985).
}

\lref\SiegelSV{
  W.~Siegel,
  ``16/16 Supergravity,''
  Class.\ Quant.\ Grav.\  {\bf 3}, L47 (1986).
}

\lref\HuangTN{
  X.~Huang and L.~Parker,
  ``Clarifying Some Remaining Questions in the Anomaly Puzzle in ${\cal N} = 1$
  Supersymmetric Yang-Mills Theory,''
  arXiv:1001.2364 [hep-th].
}

\lref\DvaliZH{
  G.~Dvali, R.~Kallosh and A.~Van Proeyen,
  ``D-term strings,''
  JHEP {\bf 0401}, 035 (2004)
  [arXiv:hep-th/0312005].
}

\lref\KachruAW{
  S.~Kachru, R.~Kallosh, A.~D.~Linde and S.~P.~Trivedi,
  ``De Sitter vacua in string theory,''
  Phys.\ Rev.\  D {\bf 68}, 046005 (2003)
  [arXiv:hep-th/0301240].
}

\lref\DouglasES{
  M.~R.~Douglas and S.~Kachru,
  ``Flux compactification,''
  Rev.\ Mod.\ Phys.\  {\bf 79}, 733 (2007)
  [arXiv:hep-th/0610102].
}

\lref\RandallUK{
  L.~Randall and R.~Sundrum,
  ``Out of this world supersymmetry breaking,''
  Nucl.\ Phys.\  B {\bf 557}, 79 (1999)
  [arXiv:hep-th/9810155].
}

\lref\NovikovUC{
  V.~A.~Novikov, M.~A.~Shifman, A.~I.~Vainshtein and V.~I.~Zakharov,
  ``Exact Gell-Mann-Low Function Of Supersymmetric Yang-Mills Theories From
  Nucl.\ Phys.\  B {\bf 229}, 381 (1983).
}

\lref\NovikovIC{
  V.~A.~Novikov, M.~A.~Shifman, A.~I.~Vainshtein and V.~I.~Zakharov,
  ``Supersymmetric instanton calculus: Gauge theories with matter,''
  Nucl.\ Phys.\  B {\bf 260}, 157 (1985)
  [Yad.\ Fiz.\  {\bf 42}, 1499 (1985)].
}

\lref\NovikovRD{
  V.~A.~Novikov, M.~A.~Shifman, A.~I.~Vainshtein and V.~I.~Zakharov,
  ``Beta Function In Supersymmetric Gauge Theories: Instantons Versus
  Traditional Approach,''
  Phys.\ Lett.\  B {\bf 166}, 329 (1986)
  [Sov.\ J.\ Nucl.\ Phys.\  {\bf 43}, 294.1986\ YAFIA,43,459 (1986\ YAFIA,43,459-464.1986)].
}

\lref\JonesIP{
  D.~R.~T.~Jones,
  ``More On The Axial Anomaly In Supersymmetric Yang-Mills Theory,''
  Phys.\ Lett.\  B {\bf 123}, 45 (1983).
}

\lref\JonesMI{
  D.~R.~T.~Jones, L.~Mezincescu and P.~C.~West,
  ``Anomalous Dimensions, Supersymmetry And The Adler-Bardeen Theorem,''
  Phys.\ Lett.\  B {\bf 151}, 219 (1985).
}

\lref\NovikovMF{
  V.~A.~Novikov, M.~A.~Shifman, A.~I.~Vainshtein and V.~I.~Zakharov,
  ``Supersymmetric Extension Of The Adler-Bardeen Theorem,''
  Phys.\ Lett.\  B {\bf 157}, 169 (1985).
}

\lref\ArkaniHamedUT{
  N.~Arkani-Hamed and H.~Murayama,
  ``Renormalization group invariance of exact results in supersymmetric  gauge
  theories,''
  Phys.\ Rev.\  D {\bf 57}, 6638 (1998)
  [arXiv:hep-th/9705189].
}

\lref\ShifmanZI{
  M.~A.~Shifman and A.~I.~Vainshtein,
  ``Solution of the Anomaly Puzzle in SUSY Gauge Theories and the Wilson
  Operator Expansion,''
  Nucl.\ Phys.\  B {\bf 277}, 456 (1986)
  [Sov.\ Phys.\ JETP {\bf 64}, 428 (1986\ ZETFA,91,723-744.1986)].
}

\lref\GatesAZ{
  S.~J.~J.~Gates and W.~Siegel,
  ``Variant Superfield Representations,''
  Nucl.\ Phys.\  B {\bf 187}, 389 (1981).
}

\lref\FerraraDH{
  S.~Ferrara, L.~Girardello, T.~Kugo and A.~Van Proeyen,
  ``Relation Between Different Auxiliary Field Formulations Of N=1 Supergravity
  Coupled To Matter,''
  Nucl.\ Phys.\  B {\bf 223}, 191 (1983).
}

\lref\AkulovCK{
  V.~P.~Akulov, D.~V.~Volkov and V.~A.~Soroka,
  ``On The General Covariant Theory Of Calibrating Poles In Superspace,''
  Theor.\ Math.\ Phys.\  {\bf 31}, 285 (1977)
  [Teor.\ Mat.\ Fiz.\  {\bf 31}, 12 (1977)].
}

\lref\GatesCZ{
  S.~J.~J.~Gates, S.~M.~Kuzenko and J.~Phillips,
  ``The off-shell (3/2,2) supermultiplets revisited,''
  Phys.\ Lett.\  B {\bf 576}, 97 (2003)
  [arXiv:hep-th/0306288].
}

\lref\SeibergQD{
  N.~Seiberg,
  ``Modifying the Sum Over Topological Sectors and Constraints on
  arXiv:1005.0002 [hep-th].
}

\lref\FradkinJQ{
  E.~S.~Fradkin and M.~A.~Vasiliev,
  ``S Matrix For Theories That Admit Closure Of The Algebra With The Aid Of
  Auxiliary Fields: The Auxiliary Fields In Supergravity,''
  Lett.\ Nuovo Cim.\  {\bf 22}, 651 (1978).
}

\Title{ } {\vbox{\centerline{Comments on Supercurrent Multiplets, }
\centerline{}
\centerline{
Supersymmetric Field Theories and Supergravity}}}
\medskip

\centerline{\it Zohar Komargodski and Nathan Seiberg}
\bigskip
\centerline{School of Natural Sciences}
\centerline{Institute for Advanced Study}
\centerline{Einstein Drive, Princeton, NJ 08540}

\smallskip

\vglue .3cm

\bigskip
\noindent
We analyze various supersymmetry multiplets containing the
supercurrent and the energy-momentum tensor.  The most widely
known such multiplet, the Ferrara-Zumino (FZ) multiplet, is not
always well-defined.  This can happen once Fayet-Iliopoulos (FI)
terms are present or when the K\"ahler form of the target space is
not exact.  We present a new multiplet $\CS_{\alpha\alphadot} $
which always exists. This understanding of the supersymmetry
current allows us to obtain new results about the possible IR
behavior of supersymmetric theories. Next, we discuss the coupling
of rigid supersymmetric theories to supergravity.  When the theory has an FZ-multiplet or it has a global $R$-symmetry the standard formalism can be used.  But when this is not the case such simple gauging is impossible.  Then, we must gauge the current $\CS_{\alpha\alphadot}$.  The resulting theory has, in addition to the graviton and the gravitino, another massless chiral superfield $\Phi$ which is essential for the consistency of the theory.  Some
of the moduli of various string models play the role of $\Phi$.  Our general considerations, which are based on the consistency of supergravity, show that such moduli cannot be easily lifted thus leading to constraints on gravity/string models.

\Date{February/2010}

\newsec{Introduction}

Supersymmetric theories\foot{Throughout this note we will focus on four-dimensional theories.  We will be using $\CN=1$ superspace, but its existence is not essential to our discussion.  We simply use it to package supersymmetry multiplets in a convenient way.} have a conserved supersymmetry current $S_{\mu\alpha}$
\eqn\conserv{\pa^\mu S_{\mu\alpha}=0~.}
It is unique up to an improvement term of the form
\eqn\genimp{S_{\mu\alpha}'=S_{\mu\alpha}+\left(\sigma_{\mu\nu}\right)
^\beta_\alpha\pa^\nu s_\beta~.}
Clearly, $S_{\mu\alpha}'$ is conserved and yields the same
supercharge $Q_\alpha$ upon integrating over a space-like
hypersurface.
The supersymmetry current $S_{\mu\alpha}$ can be embedded in a supermultiplet. This multiplet should include the conserved energy-momentum
tensor $T_{\mu\nu}$, which is also ambiguous due to a possible
improvement of the form
\eqn\emimprove{T_{\mu\nu}'=T_{\mu\nu}+\left(\eta_{\mu\nu}\pa^2-
\pa_\mu\pa_\nu\right)t~.}

The most widely known such multiplet is the Ferrara-Zumino (FZ) multiplet~\FerraraPZ, $\CJ_{\alpha\alphadot}$.  It is a real superfield\foot{
We follow the Wess and Bagger conventions~\WessCP. A vector $\ell_\mu$, is often expressed in bi-spinor notation as
\eqn\bispinor{\ell_{\alpha\dot\alpha}=-2\sigma^\mu_{\alpha\dot\alpha}\ell_\mu,\qquad
 \ell_\mu={1\over 4}\bar\sigma_\mu^{\dot\alpha\alpha}\ell_{\alpha\dot\alpha}~.}
We sometimes use
 \eqn\usefulfor{\eqalign{ & \{D_\alpha,\bar D _\alphadot
 \}= -2i\sigma^\mu _{\alpha\alphadot}
\partial_\mu \equiv i \partial_{\alpha\alphadot}~,\cr
 & [D_\alpha,\bar D^2]=2 i \bar D^\alphadot \partial_{\alpha\alphadot} ~,\cr&
 [D^2,\bar D_\alphadot]=2iD^\alpha\pa_{\alpha\alphadot}~.}}}
satisfying
\eqn\sucon{\eqalign{
 &\bar D^\alphadot \CJ_{\alpha\alphadot}=  D_\alpha X~, \cr
 &\bar D_\alphadot X =0~.}}
(The component expressions of $\CJ_{\alpha\alphadot}$ and $X$ appear below.) This multiplet includes six bosonic operators from the conserved $T_{\mu\nu}$, four bosonic operators in a (non-conserved) current $\CJ_\mu\bigr|=j_\mu$ and two bosonic operators in the complex scalar $X\bigr|=x$.  Similarly, it has twelve fermionic operators in the conserved $S_{\mu\alpha}$ and its complex conjugate.  As expected in a supersymmetric theory, the number of bosonic operators is the same as the number of fermionic operators.

The pair $(\CJ_\mu,X)$ can be transformed as
\eqn\imprFZ{\eqalign{&\CJ'_{\alpha\alphadot}=\CJ_{\alpha\alphadot}
-i\pa_{\alpha\alphadot}\left( \Xi-\bar\Xi\right)=
\CJ_{\alpha\alphadot}+[D_\alpha, \bar D_\alphadot]\left(
\Xi+\bar\Xi\right)~,\cr& X'=X+{1\over2}\bar D^2 \bar\Xi~,\cr &\bar
D_\alphadot \Xi=0~.}} This preserves the defining equation \sucon\
and acts on the components as improvement transformations like
in~\genimp\emimprove.

If $X=0$ (or more precisely, if $X= -{1\over2}\bar D ^2 \bar \Xi$ for a well-defined chiral $\Xi$), the theory is superconformal and the bottom component of $\CJ_{\alpha\dot\alpha}$ is the superconformal $R$-symmetry. In fact, the bottom component of $\CJ_{\alpha\dot\alpha}$ is conserved if and only if the theory is superconformal.

Another multiplet, which is somewhat less known, exists whenever the theory has a continuous $R$-symmetry (see e.g.\ section~7 of~\GatesNR).  We will refer to it as the $\CR$-multiplet.  Its bottom component is the conserved $U(1)_R$ current $j_\mu^{(R)}$.  It is a real superfield, $\CR_{\alpha\alphadot}$, satisfying
  \eqn\Rcon{\eqalign{
 &\bar D^\alphadot \CR_{\alpha\alphadot}=  \chi_\alpha~, \cr
 &\bar D_\alphadot \chi_\alpha = \bar D_\alphadot \bar \chi^{\alphadot} - D^\alpha \chi_\alpha =0~.}}
(The component expressions of $\CR_{\alpha\alphadot}$ and $\chi_\alpha$ appear below.) Note that $\chi_\alpha$ has the structure of a field strength
chiral superfield. Equation~\Rcon\ immediately implies $\pa^\mu\CR_\mu=0$. Therefore, $j_\mu^{(R)}$ is conserved.  Like the FZ-multiplet, this multiplet also includes twelve bosonic operators and twelve fermionic operators.

It is often the case that a theory has several continuous $R$-symmetries.  They differ by a continuous conserved non-$R$-symmetry.  The latter is characterized by a real linear superfield $J$ ($D^2 J=0$).  This ambiguity in the $\CR$-multiplet is
\eqn\globalmix{\eqalign{&\CR_{\alpha\dot\alpha}'=\CR_{\alpha\dot\alpha}
+[D_\alpha,\bar D_{\dot\alpha}]J~,\cr
& \chi_\alpha'= \chi_\alpha+{3\over 2}\bar D^2 D_\alpha J~,\cr
&D^2 J=0~.}}
It affects the supercurrent and energy-momentum tensor through improvement terms~\genimp\emimprove.

If a theory has an FZ-multiplet~\sucon, it is easy to show that it has an exact $U(1)_R$ symmetry if and only if there exists a real and well-defined\foot{Note that formally it is always possible to solve this equation with a nonlocal real operator~$U$.} $U$ such that  $\bar D^2 U=-2X$ (this normalization is for later convenience).  Intuitively, $U$ includes a non-conserved ordinary (non-$R$) current.  The equation $\bar D^2 U =-2 X $ means that the violation of its conservation is similar to that of the $R$-current at the bottom of the FZ-multiplet.  Therefore, the shift
 \eqn\RJrel{R_{\alpha\alphadot} = \CJ_{\alpha\alphadot} + [D_\alpha , \bar D_\alphadot] U}
leads to a conserved $R$-current.  Indeed, it is easy to check that this current satisfies~\Rcon\ with $\chi_\alpha = {3\over 2} \bar D^2 D_\alpha U$.

However, not every theory has such supersymmetry multiplets.  First, it is clear that if the theory does not have a continuous $R$-symmetry, $\CR_{\alpha\alphadot}$ does not exist.  It is less obvious that the FZ-multiplet $\CJ_{\alpha\alphadot}$ is not always well-defined.  It was pointed out in~\KomargodskiPC\ that when the theory has Fayet-Iliopoulos terms the FZ-multiplet is not gauge invariant.  We will show in section 2 that when the K\"ahler form of the target space is not exact the FZ-multiplet is not globally well-defined and hence does not correspond to a good operator in the theory.

This motivates us to look for another multiplet for the supersymmetry current and the energy-momentum  tensor which exists in all theories.  We propose to consider the multiplet\foot{Such a multiplet was considered in \ClarkJX, but was rejected as not having a conserved energy-momentum  tensor.  Our discussion below demonstrates that such a conserved tensor exists.} $\CS_{\alpha\dot\alpha}$ which ``interpolates'' between \sucon\ and \Rcon:
 \eqn\DSmu{
\eqalign{&\bar D^\alphadot \CS_{\alpha\alphadot} = D_\alpha X
+\chi_\alpha~,\cr &\bar D_\alphadot X =0~,\cr &\bar
D_\alphadot\chi_\alpha = \bar D_\alphadot\bar \chi^\alphadot-
D^\alpha \chi_\alpha =0~.}} We will see that this multiplet exists
for every supersymmetric theory. In some cases, if we can solve
\eqn\UXchi{\bar D^2 U=-2X ~,\qquad {\rm or} \qquad \bar D^2
D_\alpha U = -{2\over 3}\chi_\alpha} with a well-defined real $U$,
it can be explicitly improved and reduced to either~\Rcon\
or~\sucon.

In section 2 we study the multiplet~\DSmu\ in detail and clarify its relation to~\sucon\ and~\Rcon.

In section 3 we discuss the three multiplets~\DSmu\sucon\Rcon\ in simple cases and clarify when each of them exists.

In section 4 we present field-theoretic applications of our
multiplets.  We review the discussion in~\KomargodskiPC\ about
FI-terms.  We then present a similar argument for theories with
nontrivial target spaces.  In both cases we find that if the UV
theory has neither FI-terms nor non-trivial target space topology,
then it possesses an FZ-multiplet~\sucon. Therefore, the
low-energy theory must also have the same multiplet (since~\sucon\
is an operator equation). This immediately shows that FI-terms
cannot be generated (even for emergent gauge groups in the IR),
and also that the topology of the quantum moduli space of the
theory is constrained.  More explicitly, we show that starting with a renormalizable field theory without FI-terms and flowing to the IR the K\"ahler form of the quantum moduli space must be exact, and in particular, it cannot be compact (except of course isolated points).

In section~5 we couple supersymmetric field theories to supergravity.  Here we study rigid theories whose parameters are independent of $M_P$ and couple them to linearized supergravity at the leading order in $1\over M_P$.  This setup excludes theories with parameters of order the Planck scale such as FI-terms $\xi \sim M_P^2$ and nonlinear sigma models with $f_\pi \sim M_P$.

If the FZ-multiplet exists, it can be gauged; i.e.\ coupled to
supergravity. This naturally gives rise to the ``old minimal
supergravity''~\refs{\StelleYE\FerraraEM-\FradkinJQ} formalism.
Theories without FZ-multiplets (e.g.\ theories with FI-terms or
non-trivial target spaces) can still be coupled to supergravity using the old minimal formalism provided certain conditions are satisfied.  For example, this is possible when the theory has a continuous $R$-symmetry.  In this case the $\CR$-multiplet exists and we can gauge it.  The resulting theory is related to the ``new
minimal supergravity''~\refs{\AkulovCK,\SohniusTP}.  This way of constructing such theories leads to a new perspective on the construction of~\FreedmanUK\ and the results of~\refs{\BarbieriAC\KalloshVE-\DvaliZH} which were based on the ``old minimal formalism'' (see also the recent papers~\refs{\DienesTD,\KuzenkoYM}).  We review the supergravity that we obtain by gauging the $\CR$-multiplet in an appendix.  It should be emphasized that as explained in~\FerraraDH, the resulting supergravity is the same as the one obtained using the old formalism.

However, gravity theories with continuous global symmetries are
expected to be inconsistent. Therefore, we cannot base the
consistency of the theory on the existence of an exact continuous
$R$-symmetry. This leads us to the study of theories without an
$R$-symmetry and without an FZ-multiplet.  We emphasize that such theories cannot be coupled to minimal supergravity.  The simplest
possibility then is to couple the $\CS$-multiplet to supergravity.
This turns out to be related to ``16/16
supergravity''~\refs{\GirardiVQ\LangXK-\SiegelSV}.  We will limit
ourselves to the linearized theory (leading order in $1/M_p$) and
will derive the fact that in addition to the graviton and the
gravitino the theory includes a propagating chiral
``matter'' superfield $\Phi$ (or equivalently a linear multiplet).
We will study the constraints on the coupling of this superfield.
In particular, the obstruction to the existence of the
FZ-multiplet is that the equation $\bar D^2 D_\alpha U=-{2\over
3}\chi_\alpha $ \UXchi\ cannot be solved with a well-defined $U$.
The new superfield $\Phi$ couples through the combination
 \eqn\Uhatd{\hat U= U + \Phi+\Phi^\dagger}
which is well-defined.

Special cases include the relation to the absence of supergravity
theories with FI-terms~\KomargodskiPC\ and a connection with the
results of~\WittenHU\ about the quantization of Newton's
constant.

Section~6 summarizes our results.  Here we discuss aspects of moduli stabilization and use our conclusions to constrain gravity/string models, including various string constructions like D-inflation, sequestered models, flux vacua, etc.

\newsec{The $\CS$-Multiplet}

The multiplet defined below is a new option for embedding the supercurrent and energy-momentum tensor in a superfield. The advantage of this multiplet is that it exists in many examples where the others do not. Its defining properties are
 \eqn\DSmui{
\eqalign{&\bar D^\alphadot \CS_{\alpha\alphadot} = D_\alpha X
+\chi_\alpha ~,\cr &\bar D_\alphadot X =0~,\cr &\bar
D_\alphadot\chi_\alpha = \bar D_\alphadot\bar \chi^\alphadot-
D^\alpha \chi_\alpha =0~.}} Clearly, this multiplet generalizes
the FZ-multiplet~\sucon\ and the $\CR$-multiplet~\Rcon.
These special cases are obtained by setting $\chi_\alpha=0$ or
$X=0$ in \DSmui, respectively. In particular, the vector in the
bottom component of this multiplet is typically not conserved.\foot{This is reflected in two equations that follow from~\DSmui: $\bar D^2 \CS_{\alpha\alphadot}=2i\pa_{\alpha\alphadot}X$ and $\pa^{\alpha\alphadot}\CS_{\alpha\alphadot}=i\left(\bar D^2 \bar X- D^2X\right)$. The latter equation shows that if $X=0$ the bottom component is a conserved $R$-current.}

It is straightforward to work out the component expression for
these superfields. The result after a little bit of algebra is
\eqn\Sdefii{\eqalign{ \CS_\mu= &
j_\mu^{(S)}+\theta\left(S_{\mu}-{1\over \sqrt 2 }\sigma_\mu\bar \psi
\right)+\bar\theta\left(\bar S_\mu+{1\over
\sqrt2}\bar\sigma_\mu\psi\right)+{i\over 2}\theta^2\partial_\mu
x^\dagger-{i\over 2}\bar\theta^2\partial_\mu
x\cr&+(\theta\sigma^\nu\bar\theta)\left(2T_{\mu\nu}-\eta_{\mu\nu}Z+{1\over
2}\epsilon_{\mu\nu\rho\sigma}\left(\pa^{\rho}
j^{(S)\sigma}+F^{(S)\rho\sigma}\right)\right) \cr
& +\theta^2 \left({i\over 2}\pa_\rho S_\mu\sigma^\rho-{i\over
2\sqrt2}\partial_\rho\bar\psi\bar\sigma^\rho\sigma^\mu\right)
\bar\theta+\bar\theta^2\theta
\left(-{i\over 2}\sigma^\rho\pa_\rho\bar S_\mu+{i\over
2\sqrt2}\sigma^\mu\bar\sigma^\rho\pa_\rho\psi\right)\cr
&+\theta^2\bar\theta^2
\left({1\over 2}\pa_\mu\pa^\nu j_\nu^{(S)}-{1\over4}\partial^2 j^{(S)}_\mu\right)~,}}
and
\eqn\Xandchi{\eqalign{
&X=x+\sqrt 2\theta\psi+\theta^2 \left(Z+i\partial^\nu
j_\nu^{(S)}\right)~,\cr
& \chi_\alpha=-i\lambda^{(S)}_\alpha+\left(\delta_\alpha^\beta
D^{(S)}-2i\sigma^\rho\bar\sigma^\tau
F^{(S)}_{\rho\tau}\right)\theta_\beta+\theta^2\sigma_{\alpha\dot\alpha}^\nu
\pa_\nu\bar\lambda^{(S)\dot\alpha}~,}}
($x$ is the lowest component of the superfield $X$ rather than a spacetime coordinate)  satisfying the additional relations
 \eqn\addrel{\eqalign{&D^{(S)}=-4T^\mu_\mu+6Z~,\cr & \lambda^{(S)}=-2i\sigma ^\mu\bar S_\mu+3i\sqrt2 \psi~.}}
In addition, the supercurrent $S_{\mu\alpha}$ is conserved and the
energy-momentum tensor $T_{\mu\nu}$ is symmetric and conserved.

We see that the multiplet includes the $12+12$ operators in the
FZ-multiplet, as well as one Weyl fermion $\psi$, a closed two-form $F_{\mu\nu}^{(S)}$ and a real scalar $Z$.  Hence it has $16+16 $ physical operators.  These additional $4+4$ operators circumvent the no-go theorem of \ClarkJX.

From the superfield~\Sdefii\  we can find the anticommutators
 \eqn\anticoS{\eqalign{
 &\{\bar Q_{\dot\beta},S_{\mu\alpha}\}=\sigma_{\alpha\dot\beta}^\nu \left(2T_{\mu\nu} -{1\over
 2}\epsilon_{\nu\mu\rho\sigma}F^{(S)\rho\sigma} -i\eta_{\nu\mu}\pa^\rho j_\rho^{(S)} +i\pa_\nu j_\mu^{(S)}-
 {1\over 2}\epsilon_{\nu\mu\rho\sigma}\pa^{\rho}j^{(S)\sigma}\right)~, \cr
 &\{Q_{\beta},S_{\mu\alpha}\}=2i\epsilon_{\lambda\beta}\left( \sigma_{\mu\rho}\right)^{\lambda}_\alpha\pa^\rho x^\dagger~.}}
Note that these anticommutators are consistent with the
conservation equation $\partial^\mu S_{\mu\alpha}=0$.  The
standard supersymmetry algebra follows provided the fields
approach zero fast enough at spatial infinity and that $\int d^3 x
F_{ij}^{(S)}$ vanishes for all nonzero spatial $i,j$.

Given the operators $(\CS_{\alpha\alphadot}, X, \chi_\alpha)$ we can transform
 \eqn\shiftCS{\eqalign{
 &\CS_{\alpha\alphadot} \to \CS_{\alpha\alphadot} + [D_\alpha, \bar D_\alphadot]U~, \cr
 &X  \to X  +{1\over 2} \bar D^2 U ~,\cr
 &\chi_\alpha  \to \chi_\alpha  +{3\over 2}  \bar D^2 D_\alpha U~, }}
with any real superfield $U$ and preserve the defining relations
\DSmui.  This transformation shifts the energy-momentum  tensor and
the supersymmetry current by improvement terms
\eqn\STimpr{\eqalign{ &S_{\mu \alpha} \to S_{\mu \alpha} -2 i
\left(\sigma_{\mu \nu}\right)_\alpha^\beta  \partial^\nu U
\bigr|_{\theta^\beta}~, \cr
& T_{\mu\nu}\to T_{\mu\nu}+{1\over 2}
\left(\pa_\mu\pa_\nu-\eta_{\mu\nu}\pa^2\right) U\bigr| ~,}} where
$U=U\bigr|+\theta^\beta U \bigr|_{\theta^\beta} + ...$~.

We interpret the bottom component of $\CS_{\alpha\alphadot} $ as
an $R$-current which is not conserved.  The $\theta\bar \theta $
component of $U$ is an ordinary (non-$R$) current which is also
not conserved.  Hence, the transformation~\shiftCS\ shifts the
non-conserved $R$-current and yields another non-conserved
$R$-current.

We consider certain special cases:\vskip3pt
\item{1.} If we can solve $X=-{1\over 2} \bar D^2 U$ with a well-defined
(i.e.~local and gauge invariant) real operator $U$, we can transform $X$ away and find the $\CR$-multiplet~\Rcon.  Now, the bottom component of
$\CS_{\alpha\alphadot}$ is a conserved $R$-current.  Conversely,
if the theory has an exact $U(1)_R$ symmetry, the
$\CR$-multiplet~\Rcon\ exists and therefore we can solve $X=-{1\over 2} \bar
D^2 U$ in terms of a well-defined $U$. Therefore, we interpret an
$X$ which cannot be written as $\bar D^2 U$ with a real operator
$U$ as the obstruction to having an $R$-symmetry. Note that the
remaining freedom in~\shiftCS\  which preserves $X=0$ restricts $U$ to satisfy $D^2 U=0$, i.e.\ $U$ is a conserved current multiplet. This has the effect of
shifting the conserved $R$-current $j_\mu^{(R)}$ by a conserved
non-$R$-current, as we explained around~\globalmix.\vskip3pt
\item{2.}  If we can solve $\chi_\alpha= -{3\over 2} \bar D^2 D_\alpha U$
with a well-defined real operator $U$, we can transform
$\chi_\alpha$ away and find the FZ-multiplet~\sucon.
The remaining freedom which preserves $\chi_\alpha=0$ restricts $U$ to the form $\Xi + \bar\Xi$ with a
chiral $\Xi$. This is the ambiguity in the FZ-multiplet we explained around~\imprFZ. Hence we interpret a
 $\chi_\alpha$ which cannot be written as $-{3\over 2}  \bar D^2
D_\alpha U$ as the obstruction to the existence of the FZ-multiplet.\vskip3pt
\item{3.} If we can write both  $X=-{1\over 2} \bar D^2 U$ and $\chi_\alpha= -{3\over 2} \bar D^2 D_\alpha \tilde U$ but with $U \not=\tilde U$, both the FZ-multiplet and the $\CR$-multiplet exist.  In this case we can simply transform from one to the other
\eqn\RoutofFZ{\CR_{\alpha\dot\alpha}=\CJ_{\alpha\dot\alpha}+[D_\alpha,\bar
D_{\dot\alpha}](U-\tilde U)~.}
This is equivalent to the discussion around~\RJrel.\vskip3pt
\item{4.} If we can simultaneously solve $X=-{1\over 2} \bar D^2 U$ and $\chi_\alpha= -{3\over 2} \bar D^2 D_\alpha U$ (with the {\it same} $U$), we can set both $X$ and $\chi_\alpha$
 to zero.  Then the theory is superconformal.

\newsec{Examples}

\subsec{Wess-Zumino Models}

As an example, let us first discuss the general sigma model, with K\"ahler potential $K(\Phi^i,\bar\Phi^{\bar i})$ and superpotential $W(\Phi^i)$.
The expressions for $\CJ_{\alpha\alphadot}$ and $X$ are
\eqn\vecsup{\eqalign{
 &\CJ_{\alpha \alphadot} = 2g_{i \bar i}( D_\alpha  \Phi^i)(\bar D_\alphadot \bar \Phi^{\bar i})
 - {2\over 3}[D_\alpha, \bar D_{\alphadot} ] K~, \cr
 &X=4 W-{1\over3}\bar D^2K~.}}
K\"ahler transformations shift $K \to K + \Lambda +  \bar\Lambda$ with a
chiral $\Lambda$. Of course, these transformations do not affect the
physics; they change \vecsup\ by improvement transformations as in~\imprFZ\ with $\Xi = -{2\over 3 }\Lambda$. The bottom component of
$\CJ_{\alpha\alphadot}$ is an $R$-current. It includes a term
which is bilinear in fermions and a purely bosonic term. The term
bilinear in fermions is manifestly invariant under K\"ahler
transformations. The bosonic part
\eqn\Rcurrbos{j_{\mu}^{bosonic}={2i\over3}\left(\pa_\mu\phi^i
\pa_iK-\pa_\mu\bar\phi^{\bar i}\pa_{\bar i}K\right)}
is not invariant under K\"ahler transformations. This has the following geometric interpretation.  The K\"ahler form $\omega \sim
\pa_{i}\pa_{\bar j}Kd\phi^i\wedge d\bar\phi^{\bar j}$ is globally well-defined. Locally it can be expressed in terms of the K\"ahler connection $\CA \sim i\pa_i Kd\phi^i-i\pa_{\bar i} Kd\bar\phi^{\bar i}$ as $\omega= d \CA$.  Hence, we identify $j^{bosonic}_\mu$ as the pullback of $\CA$ to spacetime.

We learn that when $\omega $ is not exact, $\CA$ is not globally well-defined and hence the current $j_\mu$ is not a good operator. In this case, the whole
FZ-multiplet is not well-defined.  For example, if the target space has 2-cycles with non-vanishing integral of the K\"ahler form $\omega$, the FZ-multiplet does not exist.

A point of clarification is in order here.  If we can find a globally well-defined $\CA$ there is still freedom in performing K\"ahler transformations which affect the FZ-multiplet by improvement terms.  The global obstruction we discuss here arises only when we must cover the target space with patches with nontrivial K\"ahler transformations between them.

{\it We conclude that theories with a K\"ahler form that is not
exact do not have an FZ-multiplet.}

If the theory has a $U(1)_R$ symmetry (either spontaneously broken or not), we expect to find a globally well-defined $\CR_{\alpha\dot\alpha}$-multiplet. Let us see how this comes out.  We can use a basis where our chiral
superfields $\Phi_i$ have well-defined $R$-charges, $R_i$. The condition that
there is an $R$-symmetry implies the following two constraints
\eqn\Rsymm{\sum_iR_i\Phi^i\pa_i W=2W~,\qquad
\sum_iR_i\Phi^i\pa_iK=\sum_{\bar i}R_ i\bar \Phi^{\bar
i}\pa_{\bar i}K~.} By writing $W={1\over 2}\sum_iR_i\Phi^i\pa_i W$
and using the equations of motion \eqn\eom{\bar
D^2\pa_iK=4\pa_iW~,} we can express \eqn\genXs{X= 4 W - {1\over 3}
\bar D^2 K =  \bar D^2 \left({1\over 2}\sum _iR_i\Phi^i\pa_iK
-{1\over 3}K\right)~.} Note that ${1\over 2}\sum _iR_i\Phi^i\pa_iK
-{1\over 3}K$ is a real superfield because
of the second constraint in~\Rsymm.

Now we can perform the shift \RoutofFZ\ and obtain the
$\CR$-multiplet.  This leads to \eqn\rcursigm{\eqalign{
&\CR_{\alpha\alphadot}=2g_{i\bar j}D_\alpha\Phi^i\bar
D_{\alphadot}\bar\Phi^{\bar j}-[D_\alpha,\bar D_{\alphadot}]\sum
_i R_i\Phi^i\pa_i K~, \cr &\chi_\alpha = \bar D^2 D_\alpha\left(K
-{3\over 2}\sum _iR_i\Phi^i\pa_iK \right)~.}} These operators are
invariant under all K\"ahler transformations which preserve the
$R$-symmetry. Therefore, even if the target space has a non-exact
K\"ahler form, if the theory has an $R$-symmetry, the multiplet
$\CR_{\alpha\alphadot}$ is well-defined.  Hence, the supersymmetry
current and the energy-momentum  tensor in this $\CR$-multiplet are
good operators.

Finally, let us discuss the most general case in which the target
space has a nontrivial K\"ahler form and the theory does not have
an $R$-symmetry. Our motivation is that we would like to
eventually discuss supergravity, where exact continuous global
symmetries are expected to be forbidden.

In this case neither the FZ-multiplet nor the
$\CR$-multiplet exist, but our $\CS_{\alpha\alphadot}$ exists.
Indeed, the operators \eqn\sigmaCS{\eqalign{
&\CS_{\alpha\alphadot}= 2g_{i \bar i}( D_\alpha  \Phi^i)(\bar
D_\alphadot \bar \Phi^{ \bar i})~, \cr & X =4 W ~,\cr & \chi
_\alpha = \bar D^2 D_\alpha K~,}} are globally well-defined and
satisfy \DSmui. For example, $\CS_{\alpha\alphadot}$ depends on
the K\"ahler potential only through the K\"ahler metric, which is
invariant under K\"ahler transformations.

The bottom component of $\CS_{\alpha\alphadot}$ is an $R$-current
under which all the chiral superfields have vanishing charge. It
is not conserved unless $W=0$. If $W=0$, $\CS_{\alpha\alphadot}$
coincides with the $\CR$-multiplet~\rcursigm. If $W\not=0$, $X$
measures the violation of the divergence of $j_\mu^{(S)}$.

\subsec{Gauge Fields with FI Terms}

We now consider a theory with a $U(1)$ gauge field with an FI-term
 \eqn\FIterm{\CL=\cdots+\int d^4\theta \xi V~.}
This case is easily handled by the substitution $K \to K + \xi V$
in the expressions \vecsup,~\rcursigm,~\sigmaCS
\eqn\xisub{\eqalign{
&\CJ_{\alpha\alphadot} = ... -{2\xi\over 3}  [D_\alpha,\bar D_\alphadot] V~, \cr
&X=... -{\xi \over 3} \bar D^2 V~,\cr
&\chi_\alpha =... -4 \xi W_\alpha~.}}
From~\rcursigm,\sigmaCS\ we see that $\CR_{\alpha\alphadot}$ and $\CS_{\alpha\alphadot}$ do not have explicit $\xi$ dependence.  They depend on $\xi$ through the equations of motion.

Let us emphasize the analogy between an FI-term and nontrivial
geometry. When $\xi$ is nonzero the multiplet
$\CJ_{\alpha\alphadot} $ is not gauge invariant~\KomargodskiPC. If
the theory has nonzero $\xi$ but it has an $R$-symmetry,
$\CR_{\alpha\alphadot}$ is a good gauge invariant
operator~\refs{\RocekPC,\DienesTD,\KuzenkoYM}. However, if $\xi\not=0$ and
the theory does not have an $R$-symmetry, we must use the
multiplet $\CS_{\alpha\alphadot}$.  It includes gauge invariant and conserved
$S_{\mu \alpha}$ and $T_{\mu\nu}$.

The similarities between the situation with a nontrivial target space
and when there is a nonzero $\xi$ are easily understood by
considering a simple example.  A $U(1)$ gauge theory with $n$
chiral superfields with charge one and negative $\xi$ has as its
classical moduli space of vacua ${\bf CP}^{n-1}$. (In four dimensions
this theory is quantum mechanically anomalous, but this is
irrelevant for this reasoning).  The parameter $\xi$ controls the
size of the space. The peculiarities of the FI-term in the
microscopic description which includes the gauge field translate
to nontrivial transition functions in the macroscopic theory. Hence
$\CJ_{\alpha\alphadot} $ is not gauge invariant in the short
distance theory and it is not globally well-defined in the low-energy theory.

\newsec{Applications to Field Theory}

In the previous section we explained that
theories with non-exact K\"ahler form or with an FI-term do not have a well-defined FZ-multiplet. This fact can be used to prove some non-renormalization theorems. Let us first review the argument in~\KomargodskiPC\ for the FI-term.

A theory that has no FI-term gives rise to a well-defined FZ-multiplet satisfying the operator equation~\sucon. Since this operator is well-defined, it behaves regularly along the renormalization group flow. This immediately implies that no FI-term can be generated for the original gauge group and even for gauge groups that emerge from the dynamics. This explains why models of SUSY breaking predominantly break SUSY through $F$-terms.

We can repeat the same idea for the moduli space. In the UV, we
usually start form weakly interacting particles with canonical
kinetic terms. Therefore, the K\"ahler metric is trivial and the
FZ-multiplet exists. Since this multiplet must remain well-defined
throughout the flow, it follows that the quantum moduli space is
constrained.  It has to be such that the K\"ahler form $\omega
\sim d\CA$ is exact; i.e.\ $\CA$ is a globally well-defined. Hence
the integral $\int \omega\wedge \omega\wedge \omega\cdots $ over
any compact cycle must vanish.  In particular, this means that the
whole target space cannot be compact (it can, of course, be a set
of points).\foot{ An argument for the non-compactness of moduli
space has also been put forward by Witten, as referred to
in~\AffleckVC.  It is based on supergravity considerations and the
discussion in~\WittenHU. It is similar in spirit to our discussion
here, which is purely field theoretical.}

Let us see how this works in the case of SQCD with $N_f=N_c$.\foot{We thank E.~Witten for a useful discussion about this point.} The short distance theory is characterized by the classical moduli space
\eqn\NfNcc{\CM_c = \{M,B, \tilde B ~ |~ \det M-B\tilde B=0\}~.}
At long distance the theory flows to a theory of mesons and baryons with the quantum deformed moduli space~\SeibergBZ
\eqn\NfNc{\CM = \{M,B, \tilde B ~ |~ \det M-B\tilde B=\Lambda^{2N_c}\}~.}
We see that the topology of the moduli space changes.  However, in accordance with the general result above, the K\"ahler form of $\CM$ is exact.  In fact we can argue that even the K\"ahler potential on $\CM$ is single valued and is simply inherited from a well-defined K\"ahler potential in the embedding space parameterized by $M,B,\tilde B$.  It is instructive to consider the theory with $N_f=N_c+1$ with $N_c$ massless quarks and a single light quark.  This theory is described by a smooth K\"ahler potential for the mesons and the baryons~\SeibergBZ.  Near the origin it is approximately canonical.  Clearly, the massless modes in this theory are on the moduli space $\CM$ with a globally well-defined K\"ahler potential (which can actually be extended to the full embedding space). As we increase the mass of the light quark to infinity the K\"ahler potential changes but it remains well-defined.  In the limit of infinitely large mass this is the K\"ahler potential of the $N_f=N_c$ theory on $\CM$.

It is instructive to compare these nonrenormalization theorems to those about the FI-term.  Three approaches to these nonrenormalization theorems are possible.
\item{1.} Both nonrenormalization theorems follow from the fact that the FZ-multiplet is not well-defined.  This constrains the radiative corrections and the renormalization group flow in such theories.  In both cases it prevents us from finding a macroscopic theory with a nonzero FI-term or non-exact K\"ahler form if they are absent in the short distance theory.  This is the approach we have taken in this section.
\item{2.} The authors of~\refs{\DineTA,\WeinbergUV} followed~\SeibergVC\ and promoted all coupling constants to background fields. The inability to do this for the FI-term leads to its non-renormalization.\foot{For an earlier related approach see~\ShifmanZI.}  We can follow this approach also for the K\"ahler potential $K$.  We introduce a coupling constant $\hbar$ by replacing $K \to {1\over \hbar} K$.  If $K$ is globally well-defined, we do not need to use K\"ahler transformations as we move from patch to patch.  In this case we can trivially extend $1\over\hbar$ to a real superfield (or to a chiral plus an antichiral superfield) and find complicated higher order radiative corrections.  However, if we need to cover the target space by patches which are related to each other by K\"ahler transformations, then $1\over \hbar$ cannot be promoted to a background superfield; this would ruin the invariance of the Lagrangian under K\"ahler transformations.\foot{The situation in $\CN=2$ supersymmetry in two dimensions is a bit different.  Here both the coefficient of the FI-term and $1\over \hbar$ in the case with nontrivial geometry can be promoted to the real part of a twisted chiral superfield.  This allows us to write a supersymmetric effective action for these coupling constants.  Such an analysis leads to a simple derivation~\KS\ of the nonrenormalization theorems of~\refs{\EW,\NemeschanskyYX} about radiative corrections to the K\"ahler metric in sigma-models.}  Therefore, radiative corrections can arise only at one loop.\foot{In fact, in four dimensions these corrections are quadratically divergent and therefore ambiguous.}
\item{3.} Similar nonrenormalization theorems can be derived by weakly coupling the theory to supergravity and by using the non-existence of certain supergravity theories.  We will discuss such supergravity theories in section 5 and in the appendix.

\newsec{Coupling to Supergravity }

In this section we study the coupling to supergravity of the
various supercurrent multiplets we presented above. We are only
interested in linearized supergravity, namely the leading order in
${1\over M_p}$. This approach to supergravity is taken, for
example, in~\WeinbergCR. We begin with a review of the coupling of
the FZ-multiplet to supergravity.  We then explain the coupling of
the $\CS$-multiplet to supergravity. The case of the
$\CR$-multiplet is reviewed in the appendix.

\subsec{Gauging the FZ-Multiplet}

We start by reviewing the coupling of the FZ-multiplet to
linearized gravity.  The FZ-multiplet~\sucon\ contains a conserved
energy-momentum  tensor and supercurrent and can therefore be
coupled to supergravity. The supergravity multiplet is embedded in
a real vector superfield $H_{\alpha\dot\alpha}$. The
$\theta\bar\theta$ component of $H_{\alpha\dot\alpha}$ contains
the metric field, $h_{\mu\nu}$, a two form field $B_{\mu\nu}$, and
a real scalar. The coupling of gravity to matter is dictated at
leading order by \eqn\linear{\int d^4\theta
\CJ_{\alpha\dot\alpha}H^{\alpha\dot\alpha}~.}

We should impose gauge invariance, namely,
the invariance under coordinate transformations and local
supersymmetry transformations.
The gauge parameters are embedded in a
complex superfield $L_\alpha$, which so far obey no constraints.
We assign a transformation law to the supergravity fields of the form
\eqn\transflaw{H_{\alpha\dot\alpha}'=H_{\alpha\dot\alpha}+D_\alpha
\bar L_{\dot\alpha}-\bar D_{\dot\alpha}L_\alpha~,}
where $\bar L_{\alphadot}$ is the complex conjugate of $L_\alpha$, and thus this maintains the reality condition.

Requiring that~\linear\ be invariant under these coordinate
transformations, we get a constraint on the superfield $L_\alpha$.
Indeed, invariance requires that $0=\int d^4\theta \bar
D^{\dot\alpha} \CJ_{\alpha\dot\alpha} L_{\alpha}=\int d^4\theta X
D^\alpha L_{\alpha}$. Since $X$ is an unconstrained chiral
superfield we get the complex equation\foot{Equivalently, we can consider an unconstrained superfield $L_\alpha$ and add a compensator to cancel the variation. }
 \eqn\conserv{\bar D^2D^\alpha L_\alpha=0~.}
The analog of the
Wess-Zumino gauge is that the lowest components of $H_\mu$ vanish, i.e.\ \eqn\WZgauJ{
H_\mu\bigr|=H_\mu\bigr|_\theta=H_\mu\bigr|_{\bar \theta}= 0~, }
as well as the fact that $H_\mu\bigr|_{\theta\sigma^\nu\bar\theta} $ is symmetric in $\mu$ and $\nu$.

There is also some residual gauge freedom:\vskip3pt
\item{1.} $H_\mu\bigr|_{\theta^2}$ can be shifted by any complex divergenceless vector. This leaves only one complex degree of freedom, $\partial^\mu H_\mu\bigr|_{\theta^2}$.\vskip3pt
\item{2.} The metric field $h_{\mu\nu}$ transforms as
 \eqn\transfmetric{\delta
h_{\mu\nu}=\partial_{\nu} \xi_{\mu}+\partial_{\mu} \xi_{\nu}~,}
where $\xi_\mu$ is a real vector. \vskip3pt
\item{3.} The gravitino transforms as
\eqn\gravitinotransf{\delta \Psi_{\mu\alpha}=\pa_\mu
\omega_\alpha~.}

\noindent In this Wess-Zumino gauge the components
containing the gravitino and metric take the form
\eqn\Hmiddlei{H_\mu\bigr|_{\theta\sigma^\nu\bar\theta}=h_{\mu\nu}-\eta_{\mu\nu}
h~,}
and
\eqn\gravitinoembedding{H_{\mu}\bigr|_{\bar\theta^2\theta}=
\Psi_{\mu\alpha}+\sigma_\mu\bar\sigma^\rho\Psi_\rho~.}

The top component of $H_\mu$ is a vector field which survives in
the Wess-Zumino gauge. The bosonic off-shell degrees of freedom in
$H_\mu$ consist of the complex scalar $\partial^\mu
H_\mu\bigr|_{\theta^2}$, six real degrees of freedom in the
graviton and the four real degrees of freedom in the top component
of $H_\mu$, for a total of $12$ off-shell bosons. For the
fermions, we have only the gravitino. It has $16-4=12$ off-shell
degrees of freedom. This is the old minimal multiplet of
supergravity~\refs{\StelleYE,\FerraraEM,\FradkinJQ}. This is in
accordance with the $12$ degrees of freedom in the FZ-multiplet.

A simple consistency check is to use~\linear\ to check the leading couplings of the graviton and gravitino to matter. Recalling the formula for $\CJ_{\alpha\alphadot}$ (use~\Sdefii\ with $\chi_\alpha=0$) we find
\eqn\gravitonmatter{\CL_{graviton-matter}\sim\left(h_{\mu\nu}-
\eta_{\mu\nu}h\right)\left(T^{\mu\nu}-{1\over
3}\eta^{\mu\nu}T\right)=h_{\mu\nu}T^{\mu\nu}~,}
as expected. Similarly, for the coupling of the gravitino to matter we get
 \eqn\coupmat{\CL_{gravitino-matter}\sim
\epsilon^{\alpha\beta}\left(\Psi_{\mu\alpha}
+\sigma_\mu\bar\sigma^\rho \Psi_{\rho}\right)\left(S^\mu_\beta+{1\over
3}\sigma^\mu\bar\sigma^\rho S_\rho\right)=\Psi_{\mu\alpha}S^{\mu \alpha} ~.}
We would also like to mention that in analogy with the situation in ordinary curved space, improvements of $\CJ_{\alpha\alphadot}$ as in~\imprFZ\ shift the coupling to gravity~\linear\ by a term proportional to
$\int d^4\theta\left(\Xi+\bar\Xi \right)[D_\alpha,\bar D_\alphadot]H^{\alpha\alphadot}$.

The last ingredient is the kinetic term for the graviton and
gravitino. We begin by constructing a real superfield
$E^{FZ}_{\alpha\alphadot}$ by covariantly differentiating
$H_{\alpha\alphadot}$
\eqn\Eformulai{E^{FZ}_{\alpha\betadot}=\bar D_{\dot\tau}D^2\bar
D^{\dot\tau} H_{\alpha\betadot}+\bar D_{\dot\tau}D^2\bar
D_{\betadot}H_\alpha^{\dot\tau}+D^\gamma\bar D^2D_\alpha H_{\gamma
\betadot}-2\pa_{\alpha\betadot}\pa^{\gamma\dot\tau}H_{\gamma\dot\tau}~.}
This real expression\foot{To see that the first term in~\Eformulai\  is real one can use $D^\alpha\bar D^2 D_\alpha=\bar D_\alphadot D^2\bar D^\alphadot$.} is equivalent to a different-looking expression in~\FerraraMV.  The gauge transformations~\transflaw\ act as
\eqn\transflawe{E'^{FZ}_{\alpha\betadot}= E^{FZ}_{\alpha\betadot}+[D_\alpha,\bar D_\betadot]\left(D^2\bar D_{\alphadot}\bar L^\alphadot+\bar D^2 D^\beta L_\beta\right)~.}
Note the similarity to the improvement transformations~\imprFZ.
We see that $E^{FZ}_{\alpha\betadot}$ is invariant if~\conserv\ is imposed.

$E^{FZ}_{\alpha\betadot}$ satisfies another important algebraic equation,
\eqn\anothereq{\bar
D^{\betadot}E^{FZ}_{\alpha\betadot}=D_\alpha\left(\bar
D^2[D^\gamma,\bar D^{\dot\tau}]H_{\gamma\dot\tau}\right).} The superfield
in parenthesis is chiral. Note the similarity of~\anothereq\ to
the defining property of the FZ-multiplet
itself~\sucon. The fact that $E^{FZ}$ is invariant and satisfies an
equation identical to the supercurrent superfield guarantees that
the Lagrangian
 \eqn\gravitykinetic{\CL_{kinetic}\sim M_P^2\int
d^4\theta H^\mu E^{FZ}_\mu} is invariant. This contains in components
the linearized Einstein and Rarita-Schwinger terms. The six
additional supergauge-invariant bosons,
$\pa^\mu H_\mu\bigr|_{\theta^2}$, $H_\mu\bigr|_{\theta^4}$ are
auxiliary fields which are easily integrated out yielding
$\pa^\mu H_\mu\bigr|_{\theta^2}\sim ix$, $H_\mu\bigr|_{\theta^4}\sim j_\mu$
where $x$ and $j_\mu$ are the matter operators in the supercurrent
multiplet.

We conclude that theories which have a well-defined FZ-multiplet can be coupled to supergravity in this fashion. The coupling to supergravity adds to the original theory a propagating graviton and gravitino.

If there is no FZ-multiplet but there is an $R$-symmetry, one can still use the ill defined FZ-multiplet by slightly modifying the gauging procedure to construct a consistent supergravity theory.  Alternatively, in this case we can construct the $\CR$-multiplet and couple it to supergravity.\foot{For some comments on this case see also~\refs{\DienesTD,\KuzenkoYM}.}  For example, a free supersymmetric $U(1)$ theory with an FI-term can be coupled to supergravity in this fashion, thus reproducing the component Lagrangian of~\FreedmanUK.  This gives rise to a supergravity theory with a continuous global $R$-symmetry (unless there are no charged fields in the spectrum).  This explains in a simple fashion the results about
FI-terms~\refs{\FreedmanUK,\BarbieriAC\KalloshVE-\DvaliZH,\KomargodskiPC}. We
expect that consistent theories of quantum gravity do not have such continuous symmetries. Hence, we will not pursue theories with an exact $U(1)_R$ symmetry here, but will describe them in the appendix.

\subsec{Supergravity from the $\CS$-Multiplet}

We emphasized above that various supersymmetric field theories do
not have an FZ-multiplet and the energy-momentum tensor and the
supersymmetry current must be embedded in a larger multiplet
$S_{\alpha\dot\alpha}$. In such a case the only possible
supergravity theory is the one in which this (or a larger)
multiplet is gauged. In this section we
analyze this theory and as in the previous subsection, we limit
ourselves to the analysis of the linearized theory. We will
see that this supergravity theory is not merely a different set of
auxiliary fields, there are new on-shell modes.

We begin from the coupling to matter
 \eqn\linearii{\int d^4\theta \CS_{\alpha\alphadot} H^{\alpha\alphadot}~.}
For this to be invariant under~\transflaw, we need to impose the constraints
 \eqn\consonL{\bar D^2 D^\alpha
L_\alpha=0~,\qquad \bar D_{\dot\alpha}D^2\bar
L^{\dot\alpha}=D_\alpha \bar D^2 L^\alpha~.}
The first of them already appeared in the gauging of the FZ-multiplet~\conserv\ and the second one is shown in the appendix to arise in the gauging of the $\CR$-multiplet.  Since $L_\alpha$ is more constrained here than in the previous subsection, we will find more gauge invariant degrees of freedom.

Using an arbitrary $L_\alpha$ subject to these constraints we can choose the Wess-Zumino gauge
\eqn\WZgRm{H_\mu\bigr|=H_\mu\bigr|_\theta=H_\mu\bigr|_{\bar \theta}=0~.}
The residual gauge transformations allow us to transform $H_\mu\bigr|_{\theta^2}$ by any divergence-less vector so we remain with one complex gauge invariant operator $\pa^\mu H_\mu\bigr|_{\theta^2}$.
The transformation law
 \eqn\tranmidd{\delta H_\mu\bigr|_{\theta\sigma^\nu\bar\theta}+ \delta H_\nu\bigr|_{\theta\sigma^\mu\bar\theta}=
\pa_{\mu}\xi_{\nu}+\pa_{\nu}\xi_{\mu}~,\qquad \pa^\nu \xi_{\nu}=0~.}
This means that the trace part of this symmetric tensor is invariant under the residual symmetries and therefore, the $\theta\bar\theta$ component contains the usual graviton but also an additional invariant scalar.
The antisymmetric analog of~\tranmidd\ enjoys the usual gauge transformation for
a two-form
\eqn\transfB{\delta B_{\mu\nu}=\partial_{\nu}\omega_\nu-\pa_\nu\omega_\mu~.}
We also note that the top component of $H_\mu$ is invariant.
Thus, we see that we have 16 off-shell bosonic
degrees of freedom.  The fermion is in the $\theta^2\bar\theta$ component (and its complex conjugate). It has residual gauge symmetry \eqn\transfermni{\eqalign{&
\delta\Psi_{\mu\alpha}=i\partial_\mu \omega_\alpha~,\qquad
\sigma^\mu_{\alpha\alphadot}\pa_\mu\bar \omega^{\alphadot}=0~.}}
Since $\omega_\alpha$ satisfies the Dirac equation it cannot be used to set
any further components to zero. This is analogous to the discussion about the metric~\tranmidd.  Therefore, our theory includes a gravitino as well as an additional Weyl fermion.  Thus, we have 16 off-shell fermionic degrees of freedom.

We conclude that the theory has $16+16$ fields.  This is in accord with the $16+16$ operators in the multiplet $\CS_{\alpha\alphadot}$.  This $(16,16)$ supergravity multiplet has been recognized in the supergravity literature~\refs{\GirardiVQ,\LangXK}.\foot{For an early discussion see also~\GatesAZ.} We will explain some of its important features below and then turn to derive some consequences.

It is easy to construct a kinetic term; in fact $E_{\alpha\betadot}^{FZ}$ defined in~\Eformulai\ is still invariant because the set of transformations here is smaller than when the FZ-multiplet is gauged. However, This theory has another invariant.  It is easy to see that
\eqn\transcomm{\delta\left([D^\beta,\bar D^\betadot]H_{\beta\betadot}\right)={3\over 2}
\left(D^\gamma\bar D^2 L_\gamma+\bar D_{\dot\gamma}D^2\bar L^{\dot\gamma}\right)+{1\over2}
\left(\bar D^2 D^\gamma L_\gamma+D^2\bar D_{\dot\gamma} \bar L^{\dot\gamma}\right)}
vanishes upon imposing the constraints~\consonL. Thus, $[D^\beta,\bar D^\betadot]H_{\beta\betadot}$ is invariant. We can use this observation to write an invariant kinetic term
\eqn\kineticreduce{\int d^4\theta H^{\alpha\alphadot}[D_\alpha,\bar D_{\alphadot}][D^\beta,\bar D^\betadot]H_{\beta\betadot} =\int d^4\theta \left([D,\bar D]H\right)^2~.}

To summarize, we find that this theory admits two independent kinetic terms. Thus there is one free real parameter, $r$, and the most general kinetic term is\foot{In order not to clutter the equations we set $M_p=1$ and we suppress an overall constant in front of the Lagrangian.}
\eqn\basiscsi{\int
d^4\theta\left( H^{\alpha\alphadot}E^{FZ}_{\alpha\alphadot}+{1\over2 r}
H^{\alpha\alphadot}[D_\alpha,\bar D_\alphadot][D^\beta,\bar
D^\betadot]H_{\beta\betadot}\right)~.}

Our goal now is to identify the on-shell degrees of freedom in this theory and study their couplings to matter fields.  One possibility is to substitute the most general $H_{\alpha\alphadot}$ in~\linearii\ and~\basiscsi.  Then we can identify the auxiliary fields and integrate them out.  This is the approach we took in the previous subsection.  Alternatively, we can enlarge the gauge symmetry, relaxing either one of the two constraints~\consonL\ or both, and add compensator fields. This makes the results more transparent and hence we will follow this approach here.

In order to contrast the situation with that in the previous subsection we choose to keep the constraint~\conserv\ (the first one in~\consonL) and relax the second one by adding a chiral compensator field $\lambda_\alpha$ which transforms as
\eqn\lambdatr{\delta\lambda_\alpha={3\over 2}\bar D^2L_\alpha~.}
First, the non-invariance of the coupling to matter $\int d^4\theta H^{\alpha\alphadot}\CS_{\alpha\alphadot}$ can be corrected by a adding to the Lagrangian the term $-{1\over 6}\int d^2\theta \lambda^\alpha \chi_\alpha+c.c.$.  Next, we move to the kinetic terms~\basiscsi.
The first term is invariant, but the second term is not.  This is easily fixed by adding more terms to the Lagrangian.  We end up with the invariant Lagrangian
\eqn\comptheory{\eqalign{&\CL = \int d^4\theta
\left(H^{\alpha\alphadot}E^{FZ}_{\alpha\alphadot}+ {1\over2 r}
H^{\alpha\alphadot}[D_\alpha,\bar D_\alphadot]\left([D,\bar D]H\right)+
H^{\alpha\alphadot}\CS_{\alpha\alphadot}\right)\cr
&-\left( {1\over6}\int
d^2\theta \lambda^\alpha\chi_\alpha+c.c.\right)- {1\over r }\int
d^4\theta\biggl(\left(D^\gamma\lambda_\gamma+\bar
D_{\dot\gamma}\bar \lambda^{\dot\gamma}\right)[D,\bar
D]H-{1\over2}\left(D^\gamma\lambda_\gamma+\bar D_{\dot\gamma}\bar
\lambda^{\dot\gamma}\right)^2\biggr)~.}}
The first term in the second line corrects the non-invariance of the coupling to matter and the other two terms fix the transformation of the kinetic term~\kineticreduce.

In order to display the spectrum of~\comptheory\ we introduce $G=D^\gamma\lambda_\gamma+\bar D_{\dot\gamma}\bar \lambda^{\dot\gamma}$ which is a real linear superfield (i.e.\ it satisfies $D^2 G=0$).  We also express $\chi_\alpha=-{3\over2}\bar D^2D_\alpha U$, with a real $U$.  We should remember that this $U$ might not be well-defined; e.g.\ it might not be globally well-defined or might not be gauge invariant.  In fact, the need of gauging the $\CS$-multiplet arises precisely when this $U$ is not well-defined. The Lagrangian~\comptheory\ becomes
\eqn\comptheoryi{\eqalign{\CL=&\int d^4\theta
\left(H^{\alpha\alphadot}E^{FZ}_{\alpha\alphadot}+ {1\over2 r}
H^{\alpha\alphadot}[D_\alpha,\bar D_\alphadot][D,\bar D]H+
H^{\alpha\alphadot}\CS_{\alpha\alphadot}\right)\cr&- {1\over r}\int
d^4\theta\biggl(G\left([D,\bar D]H- r U\right)-{1\over2}G^2\biggr)~.}}

Now we can dualize $G$.  This is done by viewing it as an arbitrary real superfield and imposing the constraint $D^2G=0$ by a Lagrange multiplier term
$\int d^4\theta\left(\Phi+\Phi^\dagger\right)G$ where $\Phi$ is a chiral superfield which is invariant under the supergauge transformations subject to the constraint \conserv.  This makes it easy to integrate out $G$ using its equation of motion
\eqn\eomG{G= r^2 \left(\Phi+\Phi^\dagger\right)+r^2 U+[D,\bar D]H }
to find the Lagrangian
\eqn\comptheoi{\eqalign{&\CL= \int d^4\theta\biggl(
H^{\alpha\alphadot}E^{FZ}_{\alpha\alphadot}+\left(\Phi+\Phi^\dagger+ U\right)[D,\bar D]H-{r \over 2} \left(\Phi+\Phi^\dagger+U\right)^2+
H^{\alpha\alphadot}\CS_{\alpha\alphadot}\biggr)~.}}
In this presentation the theory looks like a standard supergravity theory based on the FZ-multiplet which is coupled to a matter system which includes the original matter as well as the chiral superfield $\Phi$.  This is consistent with the counting of degrees of freedom ($4+4$ degrees of freedom in addition to ordinary supergravity) and with the identification~\SiegelSV\ of the $16/16$ supergravity as an ordinary supergravity coupled to a chiral superfield.  Note that even though the new superfield $\Phi$ originated from the gravity multiplet, its couplings are not completely determined.  At the linear order we have freedom in the  dimensionless parameter $r$ and we expect additional freedom at higher orders.

The linear multiplet $G$ in \comptheoryi\ or equivalently the chiral superfield $\Phi$ in \comptheoi\ are easily recognized as the dilaton multiplet in string theory.  There the graviton and the gravitino are accompanied by a dilaton, a two-form field and a fermion (dilatino).  These are the degrees of freedom in $G$.  After a duality transformation this multiplet turns into a chiral superfield $\Phi$.  Furthermore, as in string models, the second term in~\comptheoi\ mixes the dilaton and the trace of the linearized graviton $h_\mu^\mu$.  Both this term and the term quadratic in $\Phi$ lead to the dilaton kinetic term.

As we mentioned above, the need for the multiplet $\CS_{\alpha\alphadot}$ arises when the operator $U$ is not a good operator in the theory.  In this case the current $\CJ_{\alpha\alphadot}$ does not exist.  The couplings in~\comptheoi\ explain how the chiral field $\Phi$ fixes this problem.  Even though $U$ is not a good operator, $\hat U= \Phi+\Phi^\dagger +U$ is a good operator.  If $U$ is not gauge invariant, $\Phi$ transforms under gauge transformations such that $\hat U$ is gauge invariant.  And if $U$ is not globally well-defined because it undergoes K\"ahler transformations, $\Phi$ has similar K\"ahler transformations such that $\hat U$ is well-defined.

The result of this discussion can be presented in two different ways.  First, as we did here, we started with a rigid theory without an FZ-multiplet and we had to gauge the $\CS$-multiplet.  This has led us to the Lagrangian~\comptheoi.  Alternatively, we could add the chiral superfield $\Phi$ to the original rigid theory such that the combined theory does have an FZ-multiplet.  Then, this new rigid theory can be coupled to standard supergravity by gauging the FZ-multiplet.

Our discussion makes it clear that if we want to couple the theory to supergravity, the additional chiral superfield $\Phi$ is not an option -- it must be added.

It is amusing to compare these conclusions with the discussion in section 4. There we used the fact that it is impossible to promote the FI-term or the coupling constant characterizing the geometry to background fields.  The coupling of such theories to gravity forces us to turn these coupling constants to fields.  However, these are not background classical fields but fluctuating dynamical fields.

\newsec{Summary and applications}

In most theories the supersymmetry current and the energy-momentum  tensor can be embedded in the familiar FZ-multiplet~\sucon.  But in a number of situations this multiplet is not a good operator in the theory.  It is either non-gauge invariant or not globally well-defined.  In this case we must use the larger multiplet $\CS_{\alpha\alphadot}$ which we analyzed in this paper.

These observations about the FZ-multiplet and the $\CS$-multiplet
allowed us to prove some non-renormalization theorems.  For
example, we have shown that starting with a renormalizable gauge
theory, the moduli space of supersymmetric vacua cannot be
compact.  Similarly, the known non-renormalization theorems of
theories with FI-terms trivially follow.

Of particular interest to us was the coupling of theories without
an FZ-multiplet to supergravity.  Here we have limited  ourselves
to supersymmetric field theories in which all dimensionful
parameters are fixed and study the limit $M_p \to \infty$.  We did not study theories in which the matter
couplings depend on $M_p$.  Since the FZ-multiplet does not exist,
we have to gauge the $\CS$-multiplet. The upshot of the analysis
of this gauging is the following.  We add to the rigid theory a
chiral superfield $\Phi$ whose couplings are such that the
combined system including $\Phi$ has an FZ-multiplet.  This
determines some but not all of the couplings of $\Phi$ to the
matter fields.  In the case of the FI-term $\Phi$ Higgses the
symmetry and in the case of nontrivial target space geometry of
the rigid theory it creates a larger total space in which the
topology is simpler.  Now that we have an FZ-multiplet we can
simply gauge it using standard supergravity techniques.  In
particular, at the linearized level the couplings of $\Phi$ depend
on only one free parameter -- the normalization of its kinetic
term.

Our results fit nicely with the many known examples of string vacua.
We see that the ubiquity of moduli in string theory is a result of low energy consistency conditions in supergravity. As we emphasized above, the chiral superfield $\Phi$ is similar to the dilaton superfield in four dimensional supersymmetric string vacua.  We often have field theory limits without an FZ-multiplet.  For example, we can have a theory on a brane with an FI-term.  The field theory limit does not have an FZ-multiplet and correspondingly, $U \sim \xi V$ is not gauge invariant. This problem is fixed, as in~\comptheoi, by coupling the matter theory to $\Phi$ which is not gauge invariant as in~\DineXK.  Similarly, we often consider field theory limits with a target space whose K\"ahler form is not exact.  This happens, for instance, on D3-branes at a point in a Calabi-Yau manifold.  If the latter is non-compact we find a supersymmetric field theory on the brane which typically does not have an FZ-multiplet because $U$ is not globally well-defined.  Coupling this system to supergravity corresponds to making $M_p$ finite.  In this case this is achieved by making the Calabi-Yau compact.  Then in addition to the graviton, various moduli of the Calabi-Yau space become dynamical.  They include fields like our $\Phi$ which couple as in~\comptheoi, thus avoiding the problems with the FZ-multiplet and making the supergravity theory consistent.

This discussion has direct implications for moduli stabilization.  It is often desirable to stabilize some moduli at energies above the supersymmetry breaking scale.  In this case we have to make sure that the resulting supergravity theory is still consistent.  In particular, it is impossible to stabilize $\Phi$ in a supersymmetric way and be left with a low energy theory without an FZ-multiplet.

For example, if the low energy theory includes a $U(1)$ gauge
field with an FI-term, this term must be $\Phi$ dependent.
Furthermore, if the mass of $\Phi$ is above the scale of
supersymmetry breaking, it must be the same as the mass of the
gauge field it Higgses.  Consequently, there is no regime in which
it is meaningful to say that there is an FI-term. Similar comments
hold for theories with a compact target space.  It is impossible
to stabilize the K\"ahler moduli while allowing moduli for the
positions of branes to remain massless without supersymmetry
breaking.\foot{This conclusion can also be obtained using the
result of~\WittenHU. Such a putative stabilization leads at
energies below the mass of $\Phi$ to an effective supergravity
theory which violates the consistency conditions in~\WittenHU.
This argument can be made in spite of the modifications
to~\WittenHU\ we found in appendix~A (and the general analysis
in~\SeibergQD).}

The comments above have applications to many popular string constructions including D-inflation, flux compactifications, and sequestering.  Some of these constructions might need to be revisited.

It would be nice to explore these ideas further, and to study in more detail specific examples in string theory. The question of moduli stabilization is crucial for understanding low energy aspects of string theory and may lead us to a better understanding of the space of vacua and SUSY breaking.  It would also be nice to find additional results using our new tools. In particular, it is conceivable that sharp statements can be made about the masses of moduli by studying the full nonlinear supergravity theory.

\vskip4cm

\centerline{\bf Acknowledgements}
We would like to thank O.~Aharony, N.~Arkani-Hamed, T.~Banks, M.~Dine, A.~Dymarsky, D.~Gaiotto, K.~Intriligator, J.~Maldacena, D.~Shih, W.~Siegel, Y.~Tachikawa and E.~Witten for useful discussions. We thank T.~Dumitrescu for very helpful comments on the manuscript. We are especially grateful to M.~Rocek for stimulating comments in the early stages of this project.
The work of ZK was supported in part by NSF grant PHY-0503584 and that of NS was supported in part by DOE grant DE-FG02-90ER40542. Any opinions, findings, and conclusions or
recommendations expressed in this material are those of the author(s) and do not necessarily reflect the views of the funding agencies.

\appendix{A}{Gauging the $\CR$-Multiplet}

Theories that have an $R$-symmetry possess the $\CR$-multiplet~\Rcon.
Since the $\CR$-multiplet contains a conserved supercurrent and an energy-momentum  tensor, we can couple it to supergravity.
 We proceed along the lines of subsection~5.1. We postulate a coupling to a real superfield $H_{\alpha\alphadot}$
\eqn\lineari{\int
d^4\theta \CR_{\alpha\dot\alpha}H^{\alpha\dot\alpha}~.} The
transformation law~\transflaw\ leaves this action invariant only
for a subset of all possible superfields $L_{\alpha}$. Using the
defining properties of the $\CR$-multiplet~\Rcon\ we immediately
get the constraint \eqn\Rconst{\bar D_{\dot\alpha}D^2\bar
L^{\dot\alpha}=D_\alpha\bar D^2 L^\alpha~.} This means that the
superfield $D_\alpha\bar D^2 L^\alpha$ is pure imaginary.

The analog of the Wess-Zumino gauge turns out to be \eqn\WZgRR{H_\mu\bigr|=H_\mu\bigr|_\theta=H_\mu\bigr|_{\bar \theta}=H_\mu\bigr|_{\theta^2}=H_\mu\bigr|_{\bar\theta^2}=0~.}
Note that in the $H_\mu\bigr|_{\theta^2}$ component there is a difference with subsection 5.1. The residual gauge transformations are\vskip3pt
\item{1.} The vector in the top component $H_{\mu}\bigr|_{\theta^4}\equiv b_\mu$ transforms like a gauge field
    \eqn\transfb{\delta b_\mu\sim \partial_\mu \omega~,}\vskip3pt
\item{2.}  $B_{\mu\nu} = H_\mu\bigr|_{\theta\sigma^\nu \bar\theta} - H_\nu\bigr|_{\theta\sigma^\mu \bar\theta}$ transforms like a usual two-form field
    \eqn\transfi{\delta B_{\mu\nu}=\partial_{\nu}\omega_\nu-\pa_\nu\omega_\mu~.}\vskip3pt
\item{3.} The graviton and gravitino are the same as in~\transfmetric,\gravitinotransf.

\noindent
Note that the transformation of the vector $b_\mu$ is consistent with the coupling~\lineari\ to the conserved current in $\CR_\mu$.

The two-form $B_{\mu\nu}$ has three off-shell degrees of freedom
and the gauge field $b_\mu$ has three degrees of freedom as well. Together with the metric we find 12 bosonic degrees of freedom. The gravitino provides the $12$ fermionic degrees of freedom. This is equivalent to the ``new minimal multiplet'' of supergravity~\refs{\AkulovCK,\SohniusTP}.

Our goal now is to construct the kinetic term for this theory. We again define a superfield $E_{\alpha\alphadot}^{R}$
\eqn\ERsupgra{E^R_{\alpha\alphadot}=i\pa_{\alpha\betadot}D^\gamma\bar D^\betadot H_{\gamma\dot\alpha}-i\pa_{\beta\alphadot}D^\beta\bar D^{\dot\gamma}H_{\alpha\dot\gamma}~.}
Replacing $D^\gamma\bar D^{\betadot}$ by $[D^\gamma, \bar D^{\betadot}]$ (and similarly for the second term), it is easy to show that the expression~\ERsupgra\ is real.

Next, we have to study its transformation law under~\transflaw. This gives
\eqn\varER{\delta E^R_{\alpha\alphadot}= {1\over2}[D_\alpha,\bar D_{\alphadot}]\left(D^\gamma\bar D^2 L_\gamma+\bar D_{\dot\gamma}D^2\bar L^{\dot\gamma}\right)~. }
Thus, $E^R_{\alpha\alphadot}$ is invariant under the restricted gauge transformations~\Rconst. Note that the object in parentheses
$\left(D^\gamma\bar D^2 L_\gamma+\bar D_{\dot\gamma}D^2\bar L^{\dot\gamma}\right)$ is a linear multiplet as in~\globalmix.

Another important relation $E^R$
satisfies is \eqn\relER{D^\alpha E^R_{\alpha\alphadot}={1\over2} D^2\bar
D_\alphadot [D^\beta,\bar D^\betadot]H_{\beta\betadot}~.} These
relations guarantee that \eqn\kineticRexact{\CL_{kinetic}\sim M_P^2\int
d^4\theta H^\mu E^R_\mu~,} is invariant.

We can now summarize the Lagrangian. We will not be careful about the coefficients since our goal is to explain the qualitative behavior. The bosonic couplings of supergarvity to matter fields follow from~\lineari\
\eqn\mattergravity{\CL_{matter-garvity}=h_{\mu\nu}T^{\mu\nu}+
\epsilon^{\mu\nu\rho\sigma}\partial_
\nu B_{\rho\sigma} \left(A^{(S)}_\mu+j_\mu^{(R)}\right)+b^\mu
j_{\mu}^{(R)}~.} We have denoted
$F_{\mu\nu}^{(S)}=\partial_{\mu} A^{(S)}_{\nu}- \partial_{\nu} A^{(S)}_{\mu}$, where $F^{(S)}_{\mu\nu}$ is the
field strength appearing in the $\CR$-multiplet.\foot{The expression in components for the $\CR$-multiplet is obtained from~\Sdefii\ by setting $X=0$.} Note that this Lagrangian is invariant under all the residual gauge transformations.
The (quadratic) kinetic term for the bosonic gravitational degrees of freedom are
\eqn\kinetic{{1\over M_P^2}\CL_{kinetic}=h\partial^2h+\left(\epsilon^{\mu\nu\rho\sigma}\partial_\nu
B_{\rho\sigma}\right)^2+b_\mu
\epsilon^{\mu\nu\rho\sigma}\partial_\nu B_{\rho\sigma}~.}
Here $h\partial^2 h$ is just a shorthand for the linearized Einstein theory.
Note the absence of $b_\mu^2$ due to gauge invariance.

We see that  $b_\mu$ is an auxiliary field that can be easily integrated out to yield
\eqn\eom{\epsilon^{\mu\nu\rho\sigma}\partial_\nu
B_{\rho\sigma}\sim j_\mu^{(R)}~.} Hence, $B_{\mu\nu}$ is an auxiliary field that is solved in terms of the
$R$-current. We conclude that both the $B$-field and the vector field $b_\mu$ are auxiliary non-propagating degrees of freedom. Thus, the coupling to supergravity via the $\CR$-multiplet does not introduce new propagating degrees of freedom beyond the graviton and gravitino.

Using~\eom\ in the action we get
\eqn\actionRgravity{\CL_{total}=h\partial^2h+h_{\mu\nu}T^{\mu\nu}+A^{(S)\mu} j_\mu^{(R)} +\cdots~.}

As an example we can consider a pure $U(1)$ gauge theory. It has an $R$-current given by $\CR_{\alpha\dot\alpha}=-{4\over g^2} W_\alpha\bar W_{\dot\alpha}$.  We Denote by $A_\mu$ the elementary gauge field in the problem. For this theory it easy to see that the operator $A^{(S)}_\mu$ in the $\CR$-multiplet is just the FI-term times the fundamental gauge field, $A_\mu^{(S)}\sim \xi A_\mu$. Hence, the effect of~\actionRgravity\ is to shift the gauge charges of all the fields in the problem by their $R$-charge (proportional to the FI-term and suppressed by the Planck scale). This reproduces the results of~\refs{\BarbieriAC\KalloshVE-\DvaliZH} which was derived in the old minimal formalism about the coupling of $R$-symmetric theories with an FI-term to supergravity.  The necessity of an exact $R$-symmetry is the root of the incompatibility of these models with a complete quantum gravity theory.

Another interesting case is the ${\bf CP}^1$ sigma model. Since
this theory has no superpotential, there is an $R$-symmetry such
that all the fields carry $R$-charge zero. It is therefore
guaranteed that this theory has a well-defined
$\CR$-multiplet~\rcursigm. As we have explained in section 3, this
multiplet is globally well-defined. It can therefore be (classically) coupled
to supergravity for any value of the radius of the
sphere.\foot{This conclusion differs from~\WittenHU, which claims
that the radius of the sphere has to be quantized in units of the
Planck scale. The difference arises because of the existence of
the $R$-symmetry. For a detailed discussion of this point  see~\SeibergQD. }

\listrefs
\end